\documentclass[twocolumn,prc,preprintnumbers,nofootinbib,superscriptaddress,showpacs,a4]{revtex4-2}
\usepackage{epsfig,graphicx,float,pst-all}
\usepackage{mathrsfs}
\usepackage[utf8]{inputenc}
\usepackage{rotating}
\usepackage{url}
\usepackage{esvect}
\usepackage{subfigure}
\usepackage{multirow}
\usepackage{amsmath}
\usepackage{amssymb}
\usepackage{amsfonts}
\usepackage{bm}
\usepackage{xcolor}
\usepackage{natbib}
\usepackage{hyperref}
\usepackage[warn]{textcomp}
\usepackage{gensymb}
\usepackage{footnote}
\usepackage{siunitx}
\usepackage{tabularx}
\usepackage{physics}
\usepackage{comment}
\usepackage{bm}
\usepackage{orcidlink}
\usepackage{booktabs} 
\usepackage{array}
\usepackage{dcolumn} %
\usepackage[normalem]{ulem}

\renewcommand{\vec}[1]{\boldsymbol{#1}}
\begin{document}
	\title{Cluster and Halo Structures of Light Nuclei within the NUCLEI-PACK Framework} 
	\author{H. M. Maridi\orcidlink{0000-0002-2210-9897}}
	\email[Corresponding author: 
	]{hasan.maridi@manchester.ac.uk}
	\affiliation{Department of Physics and Astronomy, The University of Manchester, Manchester M13 9PL, UK}
	\date{\today}

\begin{abstract}
As part of the ongoing \textsc{NUCLEI-PACK} project, this study presents a semi-classical framework for exploring the microscopic geometry of light and exotic nuclei based on optimized sphere packing of nucleons and clusters. Starting from explicit nucleon coordinates generated by the packing algorithm, the model provides direct access to charge, matter, and core–valence radii, allowing quantitative analysis of clustering and halo formation. The study covers one-nucleon halo nuclei ($^{11}$Be, $^{15}$C, $^{19}$C, and $^{8}$B) and two-nucleon halo systems ($^{6}$He, $^{11}$Li, $^{19}$B, and $^{17}$Ne). For the halo systems, the fitted geometric offset parameter~$\Delta$ exhibits an inverse correlation with the nucleon separation energy, reflecting the increasing spatial decoupling between the core and valence nucleons in weakly bound configurations. The framework also reproduces characteristic neutron–neutron separations and opening angles in Borromean nuclei and qualitatively captures the geometric arrangement of $\alpha$ clusters ($^{6}$Li, $^{7}$Li, and $^{12}$C). These results demonstrate that a simple geometric framework can effectively capture the essential features of both halo and cluster structures, providing an intuitive and computationally efficient link between nuclear geometry, binding, and experimentally observed radii.
\end{abstract}
\maketitle

\section{Introduction}	
The nuclear many-body problem remains one of the most complex and compelling challenges in modern physics. Macroscopic models such as the Liquid Drop Model provide valuable intuition~\cite{Bethe1936,Meitner1939}, while large-scale quantum calculations achieve high precision~\cite{Carlson2015}. Yet, there remains a pressing need for a computationally tractable, microscopic framework that yields transparent physical insight into nuclear geometry, particularly for exotic and weakly bound systems.

The semi-classical model \textsc{NUCLEI-PACK} constructs nuclei by explicitly assigning spatial coordinates to each proton and neutron, with configurations generated through optimized sphere-packing algorithms. This approach reflects the short-range nuclear force: strong repulsion prevents nucleons from overlapping, effectively rendering them as hard spheres at close distances. 

Conceptually, this model follows a rich lineage of geometric approaches that view nuclei as structured assemblies rather than uniform drops. Meitner and Frisch, in their explanation of fission, emphasized the analogy between nucleons and random packing of spheres~\cite{Meitner1939}. Pauling’s close-packed spheron theory envisioned nuclei as dense clusters of subunits~\cite{Pauling1965a}, while crystalline lattice models explored ordered arrangements~\cite{Coo87}. Recently, packing arguments have been applied to infer effective nucleon sizes~\cite{Kai24}, and to model nucleon enrichment through binary-mixture random close packing~\cite{Zac22,Anzivino2024}.

The \textsc{NUCLEI-PACK} project was initiated in our first paper~\cite{Mar25b}, which established the global validity of the packing concept by reproducing systematic trends in nuclear charge and matter radii and binding energies across the chart of nuclides ($1 \leq A \leq 250$). That study demonstrated that optimized geometric packing, combined with a simple energetic calibration of Coulomb and surface terms, can recover the empirical scaling of nuclear size and stability.

Building upon this foundation, the present work extends the \textsc{NUCLEI-PACK} framework to exotic light nuclei, with emphasis on the spatial geometry of halo and cluster configurations. The analysis focuses on proton–neutron distributions, core–valence separations, and the emergence of extended matter densities in weakly bound systems. Through this extension, the framework aims to provide a unified geometric interpretation of both clustering and halo formation, serving as a complementary tool to \textit{ab initio} theory and experimental programs at radioactive ion beam facilities. Thus, \textsc{NUCLEI-PACK} offers a transparent geometric bridge between microscopic structure and measurable nuclear observables.

\section{The Theoretical Model}
\subsection{Two–Cluster Packing Model for Atomic Nuclei \label{sec:two-cluster}}
I consider nuclei that can be approximated as a core $(A_1,Z_1)$ plus a valence cluster $(A_2,Z_2)$, which is particularly relevant for halo systems (e.g.\ $^{11}\mathrm{Be}\!\to\!{}^{10}\mathrm{Be}+n$) and light cluster configurations (e.g.\ $^{6}\mathrm{Li}\!\to\!\alpha+d$). This model has two stages: (i) a geometric stage that produces explicit nucleon coordinates, and (ii) an analysis stage that extracts physical observables such as radii and core–valence distances.
\subsubsection{Geometric Framework (Single Cluster)}
For a cluster with $Z$ protons and $N$ neutrons ($A=Z+N$), we treat nucleons as hard spheres with distinct radii $r_p$ and $r_n$. Let $\{\vec{p}_i\}_{i=1}^{A}$ be their coordinates and $R_{\text{pack}}$ the minimal enclosing (packing) radius. 

The binary sphere-packing problem is formulated as an optimization problem where the objective is to find an arrangement that minimizes the radius of the container sphere, $R_{\text{pack}}$. This is subject to the following constraints. First, the nucleon spheres must not overlap:
\begin{equation}
	\|\vec{p}_i - \vec{p}_j\| \geq r_i + r_j, \quad \quad 1 \leq i, j \leq A, \ i \neq j.
\end{equation}
Second, all nucleons must be contained within the minimal enclosing sphere:
\begin{equation}
	\|\vec{p}_i\| \leq R_{\text{pack}} - r_i, \quad \quad 1 \leq i \leq A, \label{eq:packing}
\end{equation}
where $r_i$ corresponds to the radius of the nucleon at position $\vec{p}_i$. The problem is solved by repeated continuous optimization from randomized initial guesses and retains the best (smallest) $R_{\text{pack}}$ and the centers $\{\vec{p}_i\}$.
\subsubsection{Two–Cluster Assembly and Centering \label{sec:two-cluster-assembly}}
Let the core and valence clusters have packing radii $R_1$ and $R_2$, and internal coordinates $\{\vec{p}^{(1)}_i\}$ and $\{\vec{p}^{(2)}_j\}$ (measured about their own centers). We place the clusters in surface contact along the $+z$ axis:
\begin{equation}
\mathbf{c}_1=\mathbf{0},\qquad
\mathbf{c}_2=(0,0,R_1+R_2).
\end{equation}
The valence cluster's coordinates are then translated by $\mathbf{c}_2$. The total center of mass (with unit nucleon masses) is
\begin{equation}
\mathbf{R}_{\rm CM}=\frac{A_1\,\mathbf{c}_1+A_2\,\mathbf{c}_2}{A_1+A_2}.
\end{equation}
All nucleon coordinates are then shifted so that the nuclear CM is at the origin:
\begin{equation}
\vec{p}'^{(1)}_i=\vec{p}^{(1)}_i-\mathbf{R}_{\rm CM},\qquad
\vec{p}'^{(2)}_j=\vec{p}^{(2)}_j+\mathbf{c}_2-\mathbf{R}_{\rm CM}.
\end{equation}
The nominal nuclear bounding radius is
\begin{equation}
R_{\rm tot}=\max\!\big(\|\mathbf{c}_1-\mathbf{R}_{\rm CM}\|+R_1,\;\|\mathbf{c}_2-\mathbf{R}_{\rm CM}\|+R_2\big).
\end{equation}
The center–to–center contact distance is $d_{\rm sep}=R_1+R_2$.
\subsubsection{Inclusion of the Spatial Offset Parameter $\Delta$}
To account for the extended spatial distribution characteristic of halo nuclei, an effective geometric parameter~$\Delta$ is introduced during the packing stage. 
After the two clusters are placed in surface contact (see Sec.~\ref{sec:two-cluster-assembly}), the center of the valence cluster is shifted outward along the core–valence axis by a distance~$\Delta$:
\begin{equation}
	\mathbf{c}_2=(0,0,R_1+R_2+\Delta).
\end{equation}
This adjustment preserves the internal packing of both clusters and does not enter directly into the subsequent rms calculations. 
Physically, $\Delta$ goes beyond a static displacement: it acts as an \emph{effective amplitude of spatial fluctuation} associated with the weak binding of the valence nucleon. 
In this interpretation, $\Delta$ represents the dynamic extension of the halo relative to the compact core, reflecting the increased spatial uncertainty and reduced confining potential in systems with low separation energies. 
By including $\Delta$, the model reproduces the experimental matter and core–valence radii of one-nucleon halo nuclei while retaining a clear geometric connection to their underlying structure.
\subsubsection{Observables}
\paragraph{Root–mean–square (RMS) operator.}
For any set of coordinates $\{\vec{p}_k\}_{k=1}^{A}$ we use the CM–subtracted second moment to define the RMS radius:
\begin{equation}
	\big\langle r^2\big\rangle_{\{\vec{p}\}}=\frac{1}{A}\sum_{k=1}^A \left\|\vec{p}_k-\frac{1}{A}\sum_{\ell=1}^A \vec{p}_\ell\right\|^2,
	\qquad
	\mathrm{RMS}=\sqrt{\big\langle r^2\big\rangle_{\{\vec{p}\}}}. \label{eq:RMS}
\end{equation}
\paragraph{Charge radius.}
Let $Z$ and $N$ be the total proton and neutron counts, and $\{\vec{p}_i^{(p)}\}_{i=1}^Z$ the proton coordinates (all nucleons measured about the nuclear CM). The point–proton contribution is 
$\langle r^2\rangle_p=\frac{1}{Z}\sum_{i=1}^{Z} \|\vec{p}_i^{(p)}\|^2$. 
The total charge radius squared is
\begin{equation}
	r_{ch}^2
	=\langle r^2\rangle_p
	+ \left\langle R_{p}^2 \right\rangle
	+ \frac{N}{Z}\,\left\langle R_{n}^2 \right\rangle
	+ \Delta_{\rm DF}
	+ 3\,\sigma_p^2. \label{eq:Rc}
\end{equation}
where  
$\left\langle R_{p}^2 \right\rangle$ and $\left\langle R_{n}^2 \right\rangle$  are the intrinsic mean–square charge radii of the proton and neutron, respectively, which account for the finite spatial extent of the individual nucleons. 
The Darwin–Foldy correction $\Delta_{\rm DF}$ includes relativistic effects from the proton’s spin–orbit coupling. These intrinsic contributions ensure that even a geometrically point–like nucleus reproduces the physical charge radius observed in scattering experiments.
I use the following physical constants from Ref. \cite{Tanihata2013} (in fm$^2$): $\left\langle R_{p}^2 \right\rangle=0.769$, $\left\langle R_{n}^2 \right\rangle=-0.1161$, $\Delta_{\rm DF}=0.033$. 
The Gaussian smearing width is taken as  $\sigma_p\approx 0.76$ fm (look at Sec. \ref{sec:two-cluster-parameters}).
\paragraph{Matter radius.}
With all nucleons’ coordinates $\{\vec{p}_k\}_{k=1}^{A}$ measured about the nuclear CM, the total matter radius squared is
\begin{align}
	r_m^2
	=\frac{1}{A}\sum_{k=1}^A \|\vec{p}_k\|^2
	&+\frac{Z\,\left\langle R_{p,\text{m}}^2 \right\rangle +N\,\left\langle R_{n,\text{m}}^2 \right\rangle}{A} \nonumber \\
	& \quad +\frac{3\,\left( Z\,\sigma_p^2 +N\,\sigma_n^2\right)}{A},
	\label{eq:Rm}
\end{align}
where $\left\langle R_{p,\text{m}}^2 \right\rangle$ and $\left\langle R_{n,\text{m}}^2 \right\rangle$ denote the intrinsic matter radii of the proton and neutron, respectively. 
I use $\left\langle R_{p,\text{m}}^2 \right\rangle=0.706$ fm$^2$, $\left\langle R_{n,\text{m}}^2 \right\rangle=0.64$ fm$^2$ \cite{Mar25b}. 
The final terms, $3\sigma_p^2$ and $3\sigma_n^2$, account for the spatial smearing of the nucleon centers due to the finite width of their wave packets. 
Although the packing geometry treats nucleons as hard spheres, their quantum motion leads to a diffuse density distribution that is approximately Gaussian in character. 
The smearing width $\sigma$ therefore provides a phenomenological correction for this quantum delocalization.
For the values of the Gaussian smearing widths, see Sec. \ref{sec:two-cluster-parameters}). 
\paragraph{Core–valence RMS distance.}
Let $\mathcal{C}$ (size $A_1$) and $\mathcal{V}$ (size $A_2$) be the sets of core and valence nucleon coordinates (about the nuclear CM). The mean pair distance is
\begin{equation}
	r_{\rm cv}^2
	=\frac{1}{A_1 A_2}\sum_{i\in\mathcal{C}}\sum_{j\in\mathcal{V}}\|\vec{p}_i-\vec{p}_j\|^2\,,
\end{equation}
\paragraph{Decomposition of the Matter Radius.}
The total mean-square matter radius of a two-cluster system can be related to the individual cluster radii and their separation by the following decomposition identity. This relationship shows how the overall size of the nucleus is determined by the intrinsic size of the clusters and their external separation.

The RMS matter radii of the core and valence clusters, $r_{m,1}$ and $r_{m,2}$, are calculated from their internal coordinates relative to their own center of mass using an analogous version of Eq.~\eqref{eq:Rm}.

The first formulation relates the total matter radius to the separation distance between the cluster centers of mass ($d_{cm}$):
\begin{equation}
	r_m^2 = \frac{A_1}{A} r_{m,1}^2 + \frac{A_2}{A} r_{m,2}^2 + \frac{A_1 A_2}{A^2} d_{cm}^2
\end{equation}
where the term $d_{cm} = \|\vec{c}_1 - \vec{c}_2\|$ is the distance between the centers of mass of the two clusters, and $A_1$, $A_2$, and $A=A_1+A_2$ are their respective mass numbers. 
It is equivalent to the expression in Ref. \cite{Wal85,Mas09,Mar25}.
This identity confirms that the total nuclear radius is a combination of the internal structure of the clusters and their external separation.

An equivalent and also valid formulation expresses the relationship in terms of the distances of each cluster's center of mass ($\mathbf{c}'_1$, $\mathbf{c}'_2$) from the total nuclear center of mass:
\begin{equation}
	r_m^2
	=\frac{A_1}{A}\,R_{m,1}^2
	+\frac{A_2}{A}\,R_{m,2}^2
	+\frac{A_1\|\mathbf{c}'_1\|^2+A_2\|\mathbf{c}'_2\|^2}{A},
\end{equation}
Both forms are mathematically equivalent and are used to validate the model. The first is often preferred for its direct physical interpretation of the separation distance.

\subsubsection{Parameters used in computations \label{sec:two-cluster-parameters} }
The hard--sphere radii assigned to the proton and neutron are chosen to preserve the same effective nuclear density as established in the global NUCLEI--PACK study~\cite{Mar25b}, while incorporating the empirical packing ratio obtained by Zaccone \textit{et~al.}~\cite{Zac22}. In Ref.~\cite{Zac22}, a ratio $r_p/r_n = 0.84$ was found to reproduce the correct neutron--to--proton packing fraction in stable nuclei. 

In Paper~I~\cite{Mar25b}, equal nucleon radii $r_p^{\mathrm{Pack}}=r_n^{\mathrm{Pack}}=1.0$~fm were adopted to reproduce the global trends of nuclear radii and binding energies. To maintain the same average nucleon volume (and hence nuclear density) in the present work, the proton radius is rescaled to $r_p^{\mathrm{Pack}}=a$, and the neutron radius to $r_n^{\mathrm{Pack}}=a/0.84$, assuming a representative isotopic composition $Z/A = 1/2.5$ and $N/A = 1.5/2.5$. Equating the average occupied volume per nucleon between the two models,
\begin{equation}
	\frac{Z a^3 + N (a/0.84)^3}{A}
	= \frac{Z (1.0)^3 + N (1.0)^3}{A},
\end{equation}
yields $a = 0.89$~fm, and hence
\begin{equation}
	r_p^{\mathrm{Pack}} = 0.89~\mathrm{fm}, 
	\qquad
	r_n^{\mathrm{Pack}} = 1.06~\mathrm{fm}.
\end{equation}

These radii are used throughout the packing calculations of this paper.

The Gaussian smearing parameters for charge and matter distributions are scaled proportionally to the respective nucleon radii, following the ratio used in Paper~I, where $\sigma_p=\sigma_n=0.85$~fm for $r_p^{\mathrm{Pack}}=r_n^{\mathrm{Pack}}=1$~fm. Thus, in the present work:
\begin{equation}
	\sigma_p = 0.85\,r_p^{\mathrm{Pack}} = 0.76~\mathrm{fm}, 
	\qquad 
	\sigma_n = 0.85\,r_n^{\mathrm{Pack}} = 0.90~\mathrm{fm}.
\end{equation}
These parameters enter Eqs.~\eqref{eq:Rc} and~\eqref{eq:Rm} through the terms $3\sigma_p^2$ and $3\sigma_n^2$, respectively.

\subsection{Three–Cluster Packing Model for Atomic Nuclei}
We extend the framework here to nuclei that can be approximated as a core $(A_1,Z_1)$ and two valence clusters $(A_2,Z_2)$ and $(A_3,Z_3)$ relevant for Borromean halo systems (e.g.\ $^{11}\mathrm{Li}\!\to\!{}^{9}\mathrm{Li}+n+n$) or light three–body cluster nuclei (e.g.\ $^{6}\mathrm{He}\!\to\!\alpha+n+n$). The model again has a geometric packing stage and an analysis stage.

\subsubsection{Geometric Framework (Three Clusters)}
Each cluster is first generated by solving its internal binary sphere–packing problem to obtain the packing radius $R_i$ and nucleon coordinates $\{\vec{p}^{(i)}_k\}$ about its own center. These are the same definitions as in the two–cluster case.

The three clusters are assembled in space by first placing them in mutual surface contact to minimize the total bounding radius, generalizing the two–body condition to a compact triangular arrangement. 
This best–packing configuration defines the most compact geometric arrangement consistent with the hard–sphere constraint. 
In halo systems, where the valence nucleons are more weakly bound, this compact geometry can later be expanded to represent the extended halo configuration, as discussed in the following subsection.

The cluster centers are denoted $\mathbf{c}_1,\mathbf{c}_2,\mathbf{c}_3$, and the nuclear center of mass is
\begin{equation}
	\mathbf{R}_{\rm CM}=\frac{A_1\mathbf{c}_1+A_2\mathbf{c}_2+A_3\mathbf{c}_3}{A_1+A_2+A_3}.
\end{equation}
All coordinates are translated so that the nuclear CM is at the origin:
\begin{equation}
	\vec{p}'^{(i)}_k=\vec{p}^{(i)}_k+\mathbf{c}_i-\mathbf{R}_{\rm CM}.
\end{equation}
We define the CM–subtracted cluster centers as
\begin{equation}
	\mathbf{c}'_i = \mathbf{c}_i - \mathbf{R}_{\rm CM},
\end{equation}
which are used in all subsequent definitions of observables. The total bounding radius of the assembled system is
\begin{equation}
	R_{\rm tot}=\max_{i=1,2,3}\big(\|\mathbf{c}'_i\|+R_i\big).
\end{equation}

\subsubsection{Inclusion of the Spatial Offset Parameter $\Delta$}
For halo configurations, an effective geometric parameter~$\Delta$ is introduced to represent the extended spatial motion of the valence nucleons relative to the compact core. 
After the three clusters are arranged in their optimal geometric configuration, the valence centers are shifted outward from the core along their respective core–valence directions:
\begin{equation}
	\mathbf{c}_2 \rightarrow \mathbf{c}_2 + \Delta\,\hat{\mathbf{r}}_{12}, 
	\qquad
	\mathbf{c}_3 \rightarrow \mathbf{c}_3 + \Delta\,\hat{\mathbf{r}}_{13},
\end{equation}
where $\hat{\mathbf{r}}_{1i}$ is the unit vector from the core center to valence cluster~$i$. 
This adjustment preserves the internal packing of each cluster and their relative angular configuration, while expanding the overall three–body geometry. 
Physically, $\Delta$ acts as an \emph{effective amplitude of spatial fluctuation} associated with the weak binding of the valence nucleons. 
Rather than being a fixed displacement, it represents the collective oscillatory extension of the halo, reflecting the increased spatial uncertainty that arises in systems with low separation energies. 
Although $\Delta$ does not enter directly into the rms formulas, it governs the degree of spatial decoupling between the core and the valence clusters, enabling the model to reproduce experimental core–valence and matter radii in Borromean halo nuclei such as $^{6}$He and $^{11}$Li.

\begin{figure}[tbh]
	\centering
	\includegraphics[width=0.48\textwidth]{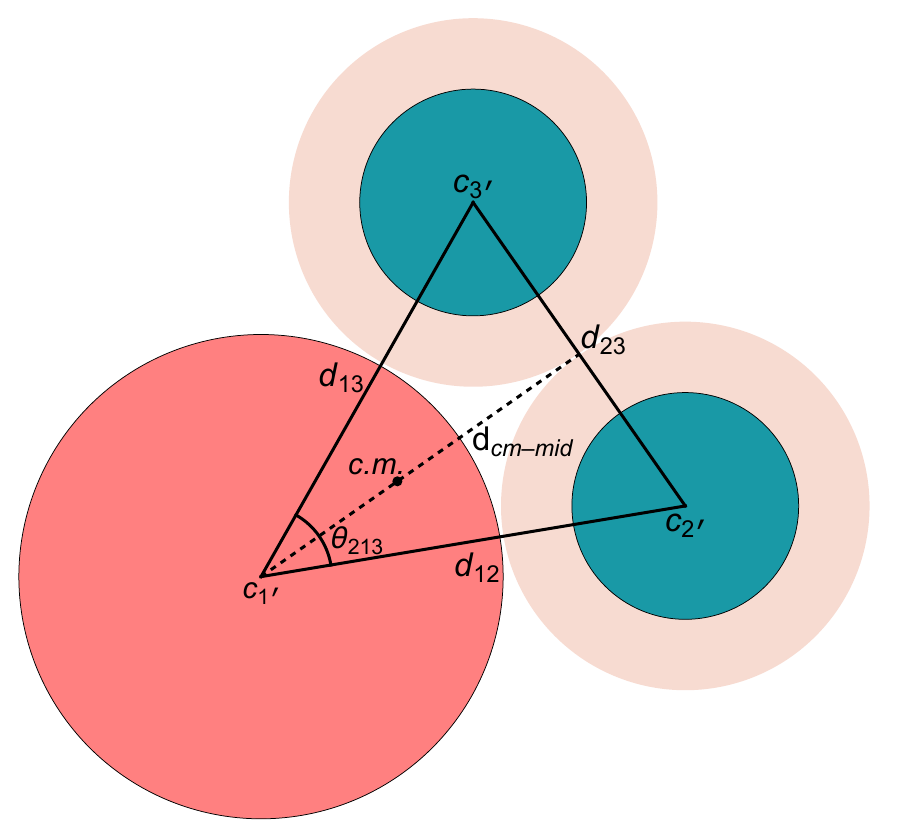}
	\caption{Schematic geometry of a three--cluster nucleus represented by a compact core (light red) and two valence clusters (bluish--green). 
		The outer reddish--orange rings illustrate the effective spatial fluctuation range associated with the offset parameter~$\Delta$. 
		The dashed line connects the core–to–midpoint distance of the valence pair, and the black point indicates the total center of mass (c.m.).}
	\label{fig:3b}
\end{figure}

Figure~\ref{fig:3b} illustrates this geometric construction within the \textsc{NUCLEI-PACK} framework. 
The centers of the core and two valence clusters, $\mathbf{c}'_{1}$, $\mathbf{c}'_{2}$, and $\mathbf{c}'_{3}$, define a triangle from which several characteristic observables are derived: the pairwise separations $d_{12}$, $d_{13}$, and $d_{23}$, the core–to–midpoint distance $d_{\mathrm{cm\!-\!mid}}$ (dashed line), and the opening angle $\theta_{213}$ at the core. 
The black point marks the total center of mass (c.m.) of the system. 
The outer shaded rings surrounding the valence clusters represent the spatial range associated with $\Delta$, which visualizes the fluctuating extension of the halo around its equilibrium geometry.

\subsubsection{Observables}
The same RMS operator as in Eq.~(\ref{eq:RMS}) is applied to any set of nucleon coordinates. The following quantities are of particular interest:

\paragraph{Charge radius.} Computed from proton coordinates relative to the nuclear CM, with intrinsic proton/neutron contributions, Darwin–Foldy term, and optional Gaussian smearing included as in Eq.~\eqref{eq:Rc}.
	
\paragraph{Matter radius.} Computed from all nucleons relative to the CM, with intrinsic nucleon matter radii and optional smearing included as in Eq.~\eqref{eq:Rm}.
	
\paragraph{Core–valence distances.} 
	\begin{enumerate}
		\item The mean distance from the core center to the midpoint of the two valence clusters:
		\begin{equation}
			d_{\mathrm{cm\!-\!mid}} = \Big\|\mathbf{c}'_1 - \frac{\mathbf{c}'_2 + \mathbf{c}'_3}{2}\Big\|.
		\end{equation}
		\item The RMS core–valence distance, defined as the mean square separation between each valence nucleon and each nucleon of the core:
		\begin{equation}
			R_{\rm cv}^2=\frac{1}{A_1(A_2+A_3)}\sum_{i\in\mathcal{C}}\sum_{j\in\mathcal{V}}\|\vec{p}_i-\vec{p}_j\|^2\,,
		\end{equation}
		where $\mathcal{C}$ and $\mathcal{V}$ denote the core and valence nucleon sets (all measured about the nuclear CM).
	\end{enumerate}
	
\paragraph{Valence–valence (neutron–neutron) distance.}
	\begin{equation}
		d_{23} = r_{nn} = \|\mathbf{c}'_2 - \mathbf{c}'_3\|.
	\end{equation}
	For single–nucleon valence clusters (e.g.\ $^{6}$He, $^{11}$Li), this is the direct neutron–neutron separation. For composite valence clusters, the same expression applies using their respective CM positions.
	
\paragraph{Opening angle at the core.}
	The three cluster centers define a triangle. The internal angles can then be extracted from scalar products of the center–to–center vectors. The opening angle at the core is 
	\begin{equation}
		\theta_{213} =
		\arccos\!\left(
		\frac{(\mathbf{c}'_2-\mathbf{c}'_1)\!\cdot\!(\mathbf{c}'_3-\mathbf{c}'_1)}
		{\|\mathbf{c}'_2-\mathbf{c}'_1\|\,\|\mathbf{c}'_3-\mathbf{c}'_1\|}
		\right).
	\end{equation}
	The angle is reported in degrees. For single–nucleon valence clusters, this corresponds directly to the neutron–neutron opening angle $\theta_{nn}$. 

\subsubsection{Decomposition of the Matter Radius}
The mean–square matter radius of the assembled three–body nucleus can be expressed as
\begin{equation}
	R_m^2 = \sum_{i=1}^{3}\frac{A_i}{A}\,R_{m,i}^2
	+ \frac{1}{A}\sum_{i=1}^{3} A_i\|\mathbf{c}'_i\|^2,
\end{equation}
where $R_{m,i}$ are the intrinsic matter radii of the clusters, $\mathbf{c}'_i$ are their CM positions relative to the total nuclear CM, and $A=A_1+A_2+A_3$. This decomposition shows that the total nuclear size arises from both the internal cluster structure and the external three–body geometry.
\subsubsection{Parameters used in computations}
The same nucleon hard–sphere radii and smearing parameters derived in the two–cluster model are adopted here, ensuring consistent nuclear density and geometric scaling across all calculations. Specifically, we use $r_p^{\mathrm{Pack}} = 0.89~\mathrm{fm}$ and $r_n^{\mathrm{Pack}} = 1.06~\mathrm{fm}$ for the packing stage, with corresponding smearing widths $\sigma_p = 0.76~\mathrm{fm}$ and $\sigma_n = 0.90~\mathrm{fm}$ as defined in Sec.~\ref{sec:two-cluster-parameters}. 

These values maintain the same mean nuclear volume established in the global \textsc{NUCLEI–PACK} calibration~\cite{Mar25b}. Calculations are presented for both best–packing geometries and $\Delta$–extended halo configurations, guided by experimental data.

\subsection{Extensions: $\alpha$ Clustering and Nuclear Dynamics}
Beyond few-body halo systems, the \textsc{NUCLEI-PACK} framework can naturally describe $\alpha$--conjugate and strongly clustered nuclei. Each $\alpha$ particle is treated as a compact four-nucleon subunit generated by the same packing procedure, allowing geometric assembly into triangular or tetrahedral configurations characteristic of light nuclei such as $^{6}$Li, $^{7}$Li, $^{12}$C and $^{16}$O. 

The same geometric principles also offer a basis for modeling collective processes such as fission and fusion. In this view, fission corresponds to the rearrangement of a packed nucleus into multiple cluster fragments connected by a neck, while fusion is the inverse process in which two pre-packed clusters merge into a single compact configuration. Such extensions demonstrate the potential of \textsc{NUCLEI-PACK} as a unified geometric language linking static structure and dynamical phenomena.

\section{Results}
\subsection{Results of the one-nucleon halo nuclei}
\begin{table*}[hbt!]
	\centering
	\small
	\setlength{\tabcolsep}{3pt}
	\renewcommand{\arraystretch}{1.1}
	\caption{\label{tab:radii1} Calculated and experimental root-mean-square (rms) charge, matter, and core–valence radii, $r_{ch}^{\mathrm{rms}}$, $r_{m}^{\mathrm{rms}}$, and $r_{cv}^{\mathrm{rms}}$, respectively, for one-nucleon halo nuclei. The parameter $\Delta$ (in fm) represents the spatial offset applied in the two-cluster packing model to reproduce the extended halo structure. Experimental values are shown for comparison where available.}
	
	\begin{tabular}{cc|c|cc|cc|cc}
		\hline\hline \noalign{\smallskip}
		\multirow{2}{*}{Nucleus}&\multirow{2}{*}{System} &\multirow{2}{*}{$\Delta$ (fm)} &
		\multicolumn{2}{c|}{$r_{ch}^{\mathrm{rms}}$ (fm)} &
		\multicolumn{2}{c|}{$r_{m}^{\mathrm{rms}}$ (fm)} &
		\multicolumn{2}{c}{$r_{cv}^{\mathrm{rms}}$ (fm)} \\
		\noalign{\smallskip} \cline{4-5}\cline{6-7}\cline{8-9} \noalign{\smallskip}
		&& & Calc. & Exp.
		& Calc. & Exp.
		& Calc. & Exp. \\
		\noalign{\smallskip} \hline  \noalign{\smallskip}	
		\multirow{3}{*}{$^{11}$Be}&$^{11}$Be
		&& 2.53& \multirow{3}{*}{\centering 2.46(2)\cite{Angeli2013} }
		& 2.55& \multirow{3}{2.0cm}{\centering 2.90\cite{Kha96}, 2.73(5)\cite{Tan88}, 2.91 \cite{Oza01a}}
		& ---& \multirow{3}{2.0cm}{\centering 5.7(4)\cite{Pal03}, 5.77(16)\cite{Fuk04}} \\
		
		&$^{10}$Be+$n$ && 2.40& & 2.64& & 4.20&  \\
		&$^{10}$Be+$n$  &1.6&2.43  &  &2.86  &  & 5.69 & \\
		\hline  \noalign{\smallskip}
		\multirow{2}{*}{$^{15}$C}&$^{15}$C
		&& 2.56& \multirow{2}{*}{\centering  }
		& 2.67& \multirow{2}{2.0cm}{\centering 2.61 \cite{Dob21}, 2.78(9)\cite{Lia90}}
		& ---& \multirow{2}{2.0cm}{\centering 4.5(2)\cite{Nak09}} \\
		
		&$^{14}$C+$n$ && 2.49& & 2.75& & 4.54&  \\
		\hline  \noalign{\smallskip}
		\multirow{3}{*}{$^{19}$C}&$^{19}$C
		&& 2.92& \multirow{3}{*}{\centering  }
		& 2.88& \multirow{3}{2.0cm}{\centering 3.01 \cite{Nak99}, 3.13(7)\cite{Oza01a}}
		& ---& \multirow{3}{2.0cm}{\centering  5.5(3)\cite{Nak99}}  \\
		
		&$^{18}$C+$n$ && 2.98& & 2.98& & 5.21&  \\
		&$^{18}$C+$n$ &0.6& 2.95& & 3.01& & 5.61&  \\
		\hline  \noalign{\smallskip}
		\multirow{3}{*}{$^{8}$B}&$^{8}$B
		&& 2.28& \multirow{3}{*}{\centering 2.89(9)\cite{Dob19}}
		& 2.28& \multirow{3}{2.0cm}{\centering 2.38(4)\cite{Tan88}, 2.58(6)\cite{Dob19}}
		& ---& \multirow{3}{2.0cm}{\centering } \\
		
		&$^{7}$Be+$p$ && 2.45& & 2.41& & 3.55&  \\
		&$^{7}$Be+$p$ &1.2& 2.76& & 2.62& & 4.71&  \\
		\hline\hline
	\end{tabular}
\end{table*}

\begin{figure*}[ht!]
	\centering
	\includegraphics[width=0.32\textwidth]{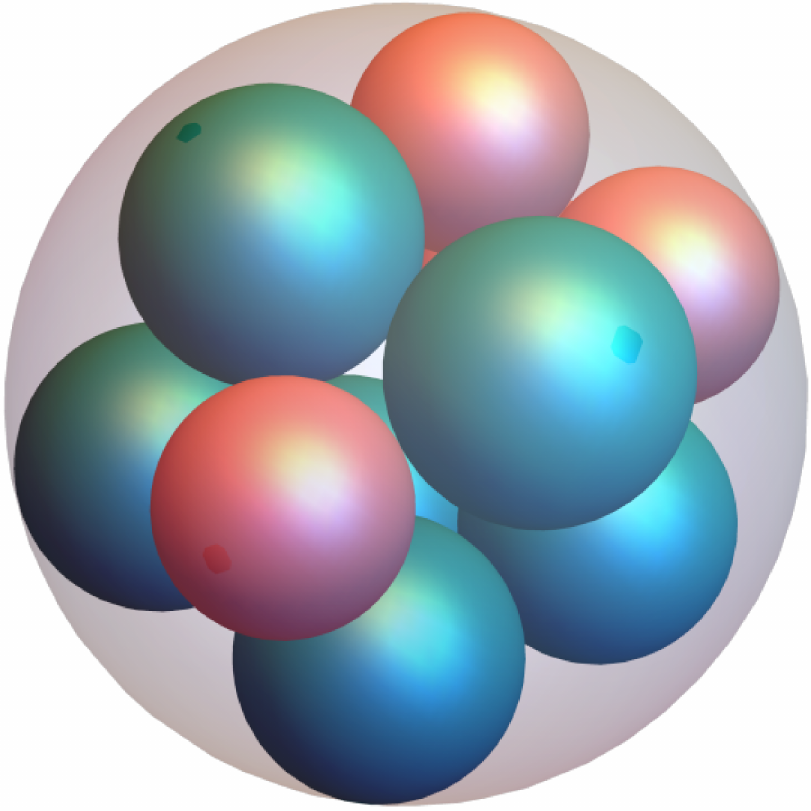}
	\includegraphics[width=0.32\textwidth]{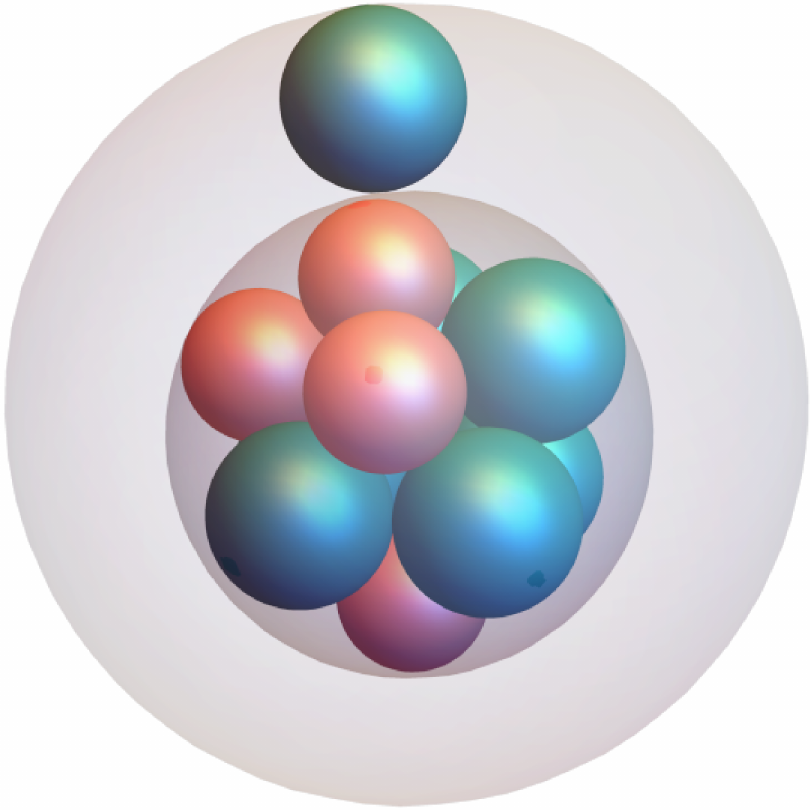}
	\includegraphics[width=0.32\textwidth]{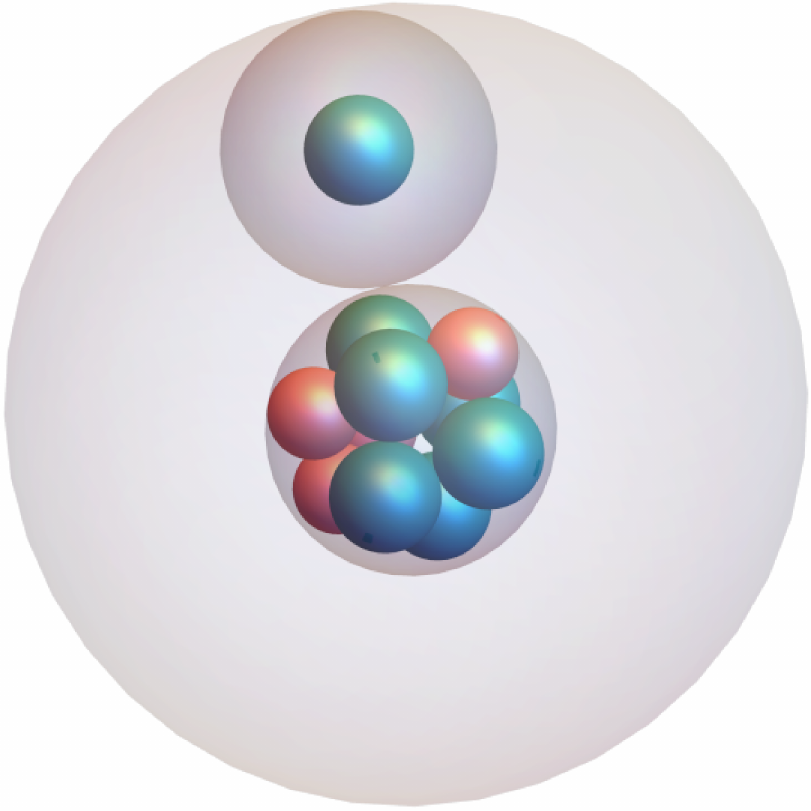}
	\caption{Visualizations of the two-cluster packing model for the one-neutron halo nucleus $^{11}$Be in three configurations: (a) as a whole nucleus, (b) as a $^{10}$Be core plus a valence neutron ($^{10}$Be + $n$), and (c) as $^{10}$Be + $n$ with an added spatial offset $\Delta$ (“delta cloud”) representing the extended halo component. Protons are shown in light red and neutrons in bluish-green.}
	\label{fig:11Be}
\end{figure*}

\begin{figure*}[ht!]
	\centering
	\includegraphics[width=0.32\textwidth]{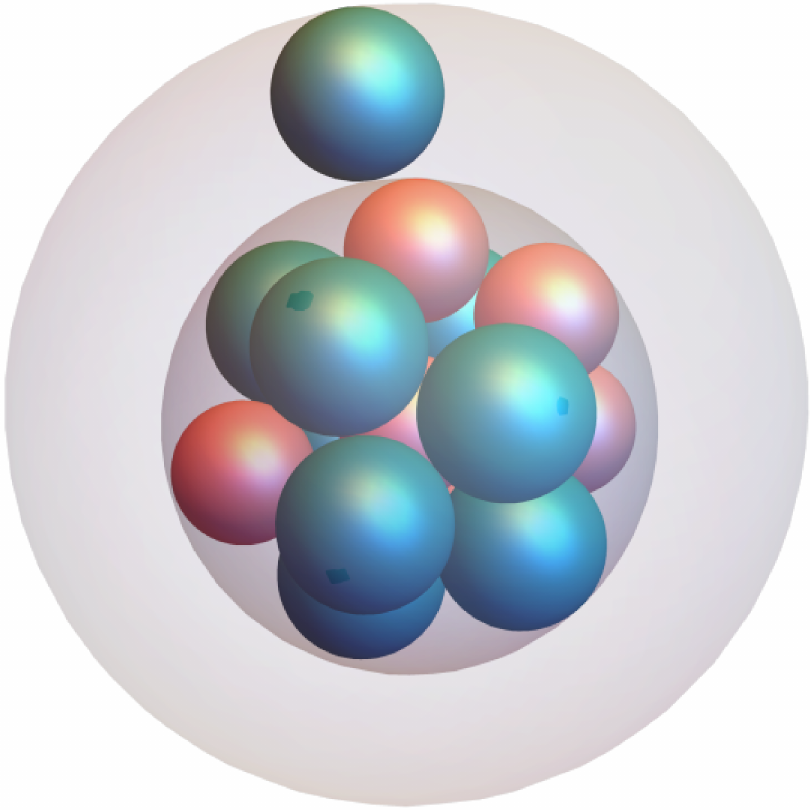}
	\includegraphics[width=0.32\textwidth]{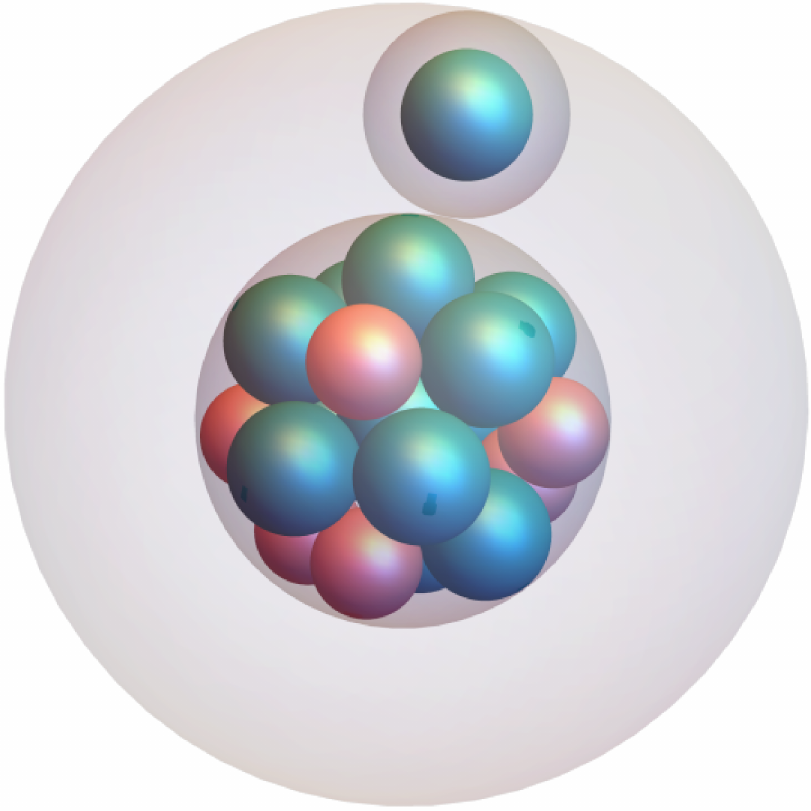}
	\includegraphics[width=0.32\textwidth]{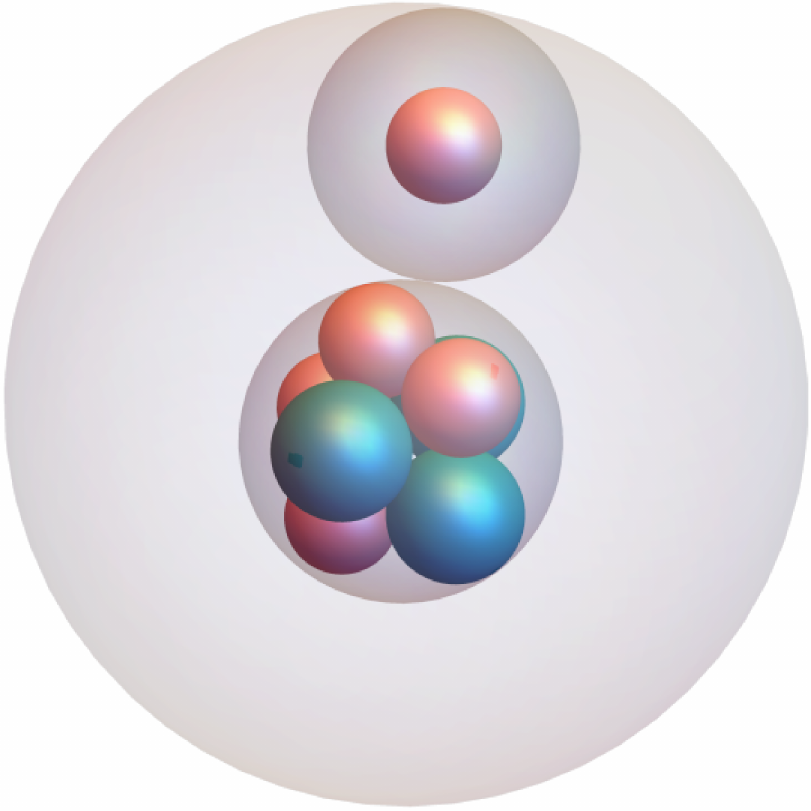}
	\caption{Visualizations of the two-cluster packing model for selected one-nucleon halo nuclei:  (a) the one-neutron halo nucleus $^{15}$C represented as $^{14}$C+$n$ (no additional spatial offset required for fitting radii), (b) $^{19}$C as $^{18}$C+$n$ with an added offset $\Delta$, and (c) the one-proton halo $^{8}$B as $^{7}$B+$p$ including the $\Delta$ extension to reproduce the observed halo size.}
	\label{fig:two-cluster-packing}
\end{figure*}

Figure~\ref{fig:11Be} shows representative packing configurations for the one-neutron halo nucleus $^{11}$Be in three cases: (a) as a whole nucleus, (b) as $^{10}$Be + $n$, and (c) as $^{10}$Be + $n$ with adding a delta Cloud.

The two-cluster packing framework provides direct insight into the microscopic structure of weakly-bound and halo nuclei. By explicitly separating the nucleus into a core and one or more valence nucleons (or clusters), the model naturally captures the spatial extent characteristic of these systems.

Table~\ref{tab:radii1} summarizes the calculated and experimental root-mean-square (rms) charge, matter, and core--valence radii for the investigated one-nucleon halo nuclei, while representative geometric configurations are illustrated in Figs.~\ref{fig:11Be} and~\ref{fig:two-cluster-packing}. In this approach, each nucleus is described as a compact core surrounded by a weakly bound valence nucleon, and the spatial offset parameter~$\Delta$ controls the degree of halo extension required to reproduce the experimental rms core--valnce radii.  For consistency across all visualizations, protons are always presented in light red and neutrons in bluish-green.

For the one-neutron halo nucleus $^{11}$Be, the calculated charge and matter radii for the unshifted two-cluster configuration ($^{10}$Be+$n$) are $r_{ch}^{\mathrm{rms}}=2.40$~fm and $r_{m}^{\mathrm{rms}}=2.64$~fm, respectively. Introducing a modest offset of $\Delta=1.6$~fm gives the matter radius of 2.86~fm, in excellent agreement with the experimental values of 2.73--2.91~fm~\cite{Tan88,Oza01a,Kha96}, while the corresponding core--valence distance $r_{cv}^{\mathrm{rms}}=5.69$~fm matches the measured range of 5.7--5.77~fm~\cite{Pal03,Fuk04}. This confirms that a small geometric displacement of the valence neutron is sufficient to reproduce the halo tail characteristic of $^{11}$Be.

The nucleus $^{15}$C exhibits a less pronounced neutron halo. A satisfactory description of the matter radius ($r_{m}^{\mathrm{rms}}=2.67$~fm) is obtained without requiring any offset ($\Delta=0$), consistent with the experimental range 2.61--2.78~fm~\cite{Lia90,Dob21}. The deduced core--valence radius $r_{cv}^{\mathrm{rms}}=4.54$~fm agrees with the experimental value of 4.5(2)~fm~\cite{Nak09}, confirming that $^{15}$C lies near the boundary between weakly-bound and halo-like structures.

In contrast, $^{19}$C requires a finite offset ($\Delta=0.6$~fm) to reproduce its strongly extended matter distribution. With this configuration, the calculated $r_{m}^{\mathrm{rms}}=3.01$~fm aligns closely with experimental measurements of 3.01--3.13~fm~\cite{Nak99,Oza01a}, and the model yields $r_{cv}^{\mathrm{rms}}=5.61$~fm, consistent with the data of 5.5(3)~fm. 

For the one-proton halo nucleus $^{8}$B, a larger offset ($\Delta=1.2$~fm) is required to match the observed charge and matter radii. The model reproduces $r_{ch}^{\mathrm{rms}}=2.76$~fm and $r_{m}^{\mathrm{rms}}=2.62$~fm, in reasonable agreement with the experimental values $r_{ch}=2.89(9)$~fm and $r_{m}=2.58(6)$~fm~\cite{Dob19}. 
The corresponding core--valence radius $r_{cv}^{\mathrm{rms}}=4.71$~fm is consistent with the value expected from the halo radius $r_{h}=4.25$~fm reported in Ref.~\cite{Dob19}, corresponding to a core–valence separation of about 4.8~fm. This agreement supports the interpretation of $^{8}$B as a well-developed one-proton halo nucleus.

Overall, the two-cluster packing model successfully reproduces the systematic trends of the experimental charge and matter radii across the one-nucleon halo nuclei. The fitted $\Delta$ values exhibit an inverse correlation with the one-nucleon separation energy, reflecting the increasing spatial separation between the core and the valence nucleon in halo systems.
The close agreement between calculated and experimental rms radii demonstrates that a simple geometric offset within the packing framework effectively captures the essential geometry of one-nucleon halo formation.

\subsection{Results of the two-nucleon halo nuclei}

\begin{table*}[tbh]
	\centering
	\small
	\setlength{\tabcolsep}{3pt}
	\renewcommand{\arraystretch}{1.1}
	\caption{\label{tab:radii2} Calculated and experimental root-mean-square (rms) charge, matter, and core–valence radii, $r_{ch}^{\mathrm{rms}}$, $r_{m}^{\mathrm{rms}}$, and $r_{cv}^{\mathrm{rms}}$, together with the neutron–neutron separation $r_{nn}$ and opening angle $\theta_{nn}$, for the studied two-nucleon halo nuclei. The parameter  $\Delta$ (in fm) denotes the spatial offset introduced in the three-cluster packing model to reproduce the extended halo configuration. Experimental values are included for comparison where available.} 
	\begin{tabular}{cc|c|cc|cc|cc|cc|cc}
		\hline\hline \noalign{\smallskip}
		\multirow{2}{*}{Nucleus}&\multirow{2}{*}{System} &
		\multirow{2}{*}{$\Delta$ (fm)}&
		\multicolumn{2}{c|}{$r_{ch}^{\mathrm{rms}}$ (fm)} &
		\multicolumn{2}{c|}{$r_{m}^{\mathrm{rms}}$ (fm)} &
		\multicolumn{2}{c|}{$r_{cv}^{\mathrm{rms}}$ (fm)}&
		\multicolumn{2}{c|}{$r_{nn}$ (fm)}&
		\multicolumn{2}{c}{$\theta_{nn}$ ($^o$)} \\
		\noalign{\smallskip} \cline{4-5}\cline{6-7}\cline{8-9} \cline{10-11}\cline{12-13}\noalign{\smallskip}
		&& & Calc. & Exp.
		& Calc. & Exp.
		& Calc. & Exp.
		& Calc. & Exp.
		& Calc. & Exp. \\
		\noalign{\smallskip} \hline  \noalign{\smallskip}	
		\multirow{4}{*}{$^{6}$He}&$^{6}$He 
		&& 2.18& \multirow{4}{1.5cm}{\centering 2.07(1)\cite{Angeli2013}} 
		& 2.23& \multirow{4}{1.55cm}{\centering 2.48(3)\cite{Tan88}, 2.67\cite{Ber07}}
		& ---& \multirow{4}{1.62cm}{\centering 3.36(9)\cite{Aum99}, 3.9(2)\cite{Sun21}} &&\multirow{4}{1.55cm}{\centering 4.1(7)\cite{Sun21}, 3.75(93)\cite{Hag07}}&&\multirow{4}{1.4cm}{\centering 56$^o$\cite{Sun21}, 52$^o$\cite{Hag07}} \\
		&$^{4}$He+$2n$ && 2.00& & 2.83& & 4.42&  & & &  & \\
		&$^{4}$He+$n$+$n$ && 2.04& & 2.49& & 3.37&  &2.12&  &38$^o$ & \\
		&$^{4}$He+$n$+$n$  &0.6&2.00  &  &2.68  &  & 3.90 & &3.32&&51$^o$& \\
		\hline  \noalign{\smallskip}
		\multirow{4}{*}{$^{11}$Li}&$^{11}$Li
		&& 2.54& \multirow{4}{1.5cm}{\centering 2.48(4)\cite{Angeli2013} }
		& 2.57& \multirow{4}{1.55cm}{\centering 3.12(6)\cite{Tan88}, 2.78(7)\cite{Lia90}}
		& ---& \multirow{4}{1.62cm}{\centering 5.01(32)\cite{Nak06}, 5.15(33)\cite{Hag07}} &&\multirow{4}{1.55cm}{\centering 5.5(2.2)\cite{Hag07}}&&\multirow{4}{1.4cm}{\centering48$^o$\cite{Nak06}, 56$^o$\cite{Hag07}} \\
		&$^{9}$Li+$2n$ && 2.51& & 2.99& & 5.23&  &&&& \\
		&$^{9}$Li+$n$+$n$  && 2.44& & 2.73& & 4.14&  &2.12&&33$^o$& \\
		&$^{9}$Li+$n$+$n$ &1.1& 2.48& & 3.00& & 5.19& &4.38&&53$^o$& \\
		\hline  \noalign{\smallskip}
		\multirow{4}{*}{$^{19}$B}&$^{19}$B
		&& 2.94& \multirow{4}{1.5cm}{  }
		& 2.90& \multirow{4}{1.55cm}{ 3.11(3)\cite{Suz99}}
		& ---& \multirow{4}{1.62cm}{ 5.75(11)\cite{Coo20}}  &&\multirow{4}{1.55cm}{\centering }&&\multirow{4}{1.4cm}{\centering 25$^o$\cite{Coo20}}  \\
		&$^{17}$B+$2n$ &&2.98 & &3.21 & &6.02 &  & & &  & \\
		&$^{17}$B+$n$+$n$ && 2.86& & 3.02& & 4.89&  &2.12&&28$^o$& \\
		&$^{17}$B+$n$+$n$ &0.9& 2.86& & 3.14& & 5.70&  &3.92&&44$^o$&\\
		\hline  \noalign{\smallskip}
		\multirow{2}{*}{$^{17}$Ne}&$^{17}$Ne
		&&2.72 & \multirow{2}{1.5cm}{\centering 3.04(1)\cite{Angeli2013} }
		&2.67 & \multirow{2}{1.55cm}{\centering 2.75(7)\cite{Tan88}}
		& ---& \multirow{2}{1.62cm}{ } &&&&\multirow{2}{1.4cm}{\centering } \\  
		
		&$^{15}$O+$p$+$p$ && 2.84& & 2.80& & 4.34&  &1.78&&26$^o$&  \\
		\hline\hline
	\end{tabular}
\end{table*}

\begin{figure*}[ht!]
	\centering
	\includegraphics[width=0.24\textwidth]{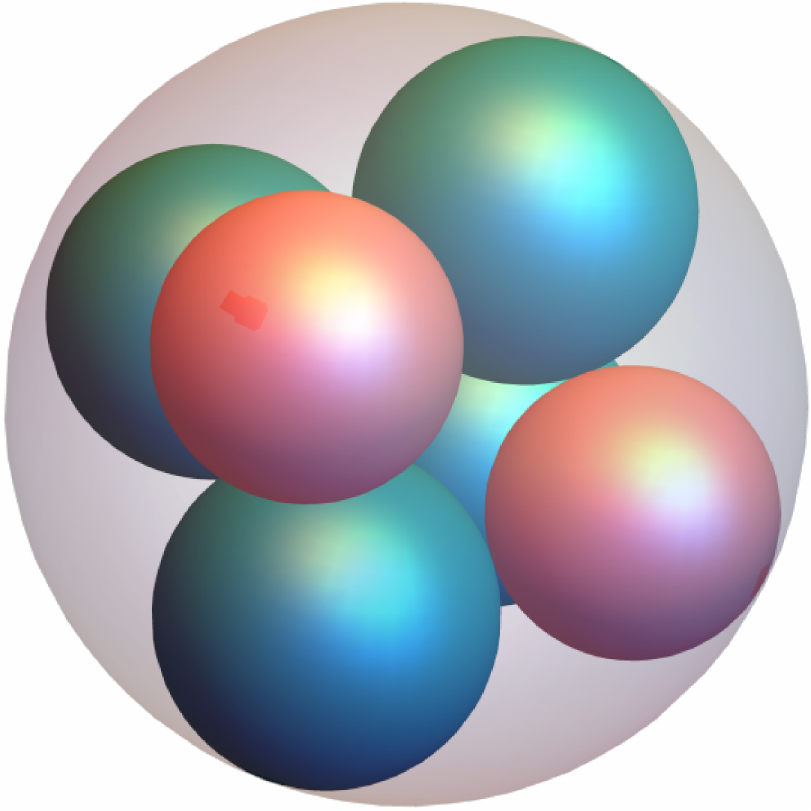}
	\includegraphics[width=0.24\textwidth]{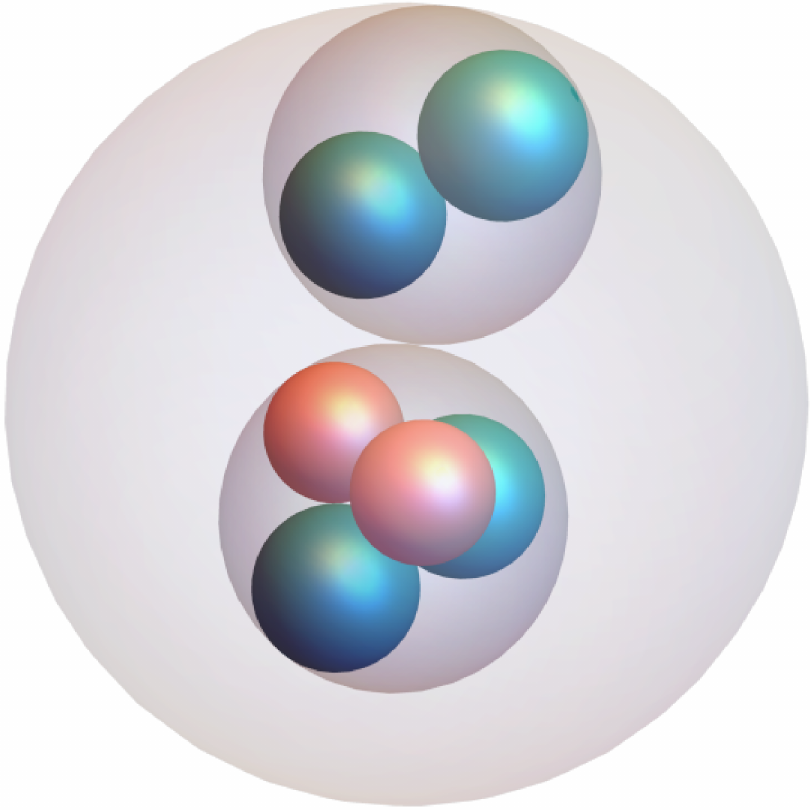}
	\includegraphics[width=0.24\textwidth]{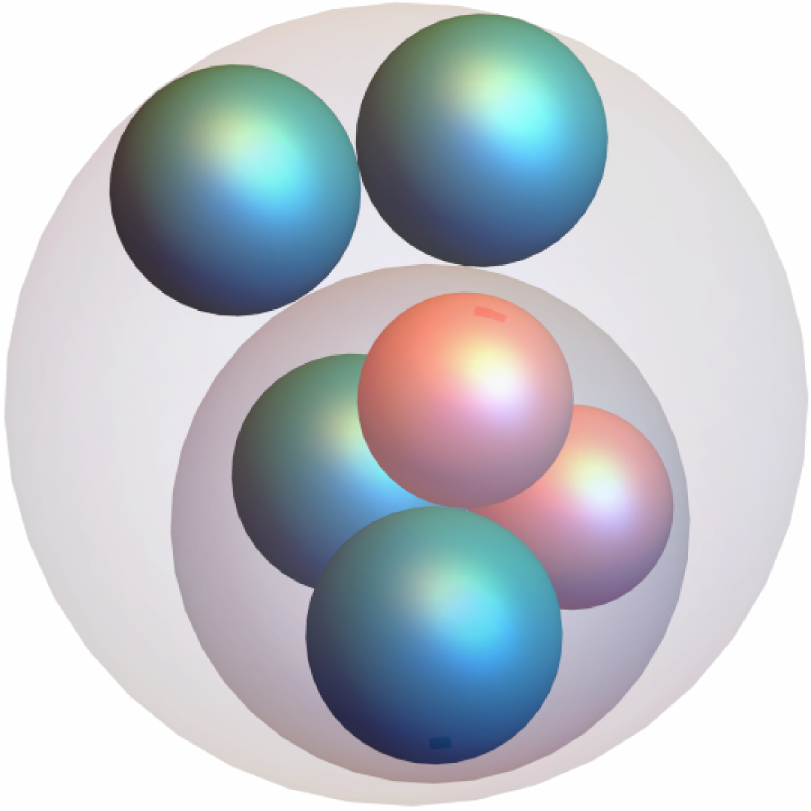}
	\includegraphics[width=0.24\textwidth]{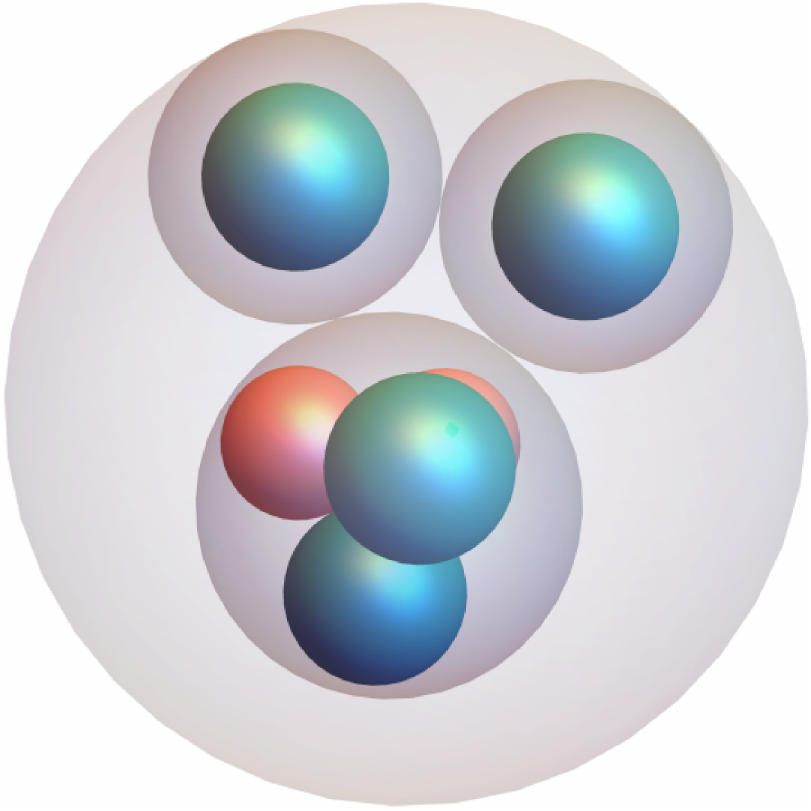}
	\caption{Visualizations of the three-cluster packing model for the two-neutron halo nucleus $^{6}$He in four configurations: (a) the whole nucleus, (b) as a $^{4}$He core plus a dineutron cluster $^{4}$He+$2n$, (c) as three-body $^{4}$He+$n$+$n$, and (d) as $^{4}$He+$n$+$n$ with an added spatial offset $\Delta$ (“delta cloud”) representing the extended halo distribution.}
	\label{fig:6He}
\end{figure*}

\begin{figure*}[ht!]
	\centering
	\includegraphics[width=0.24\textwidth]{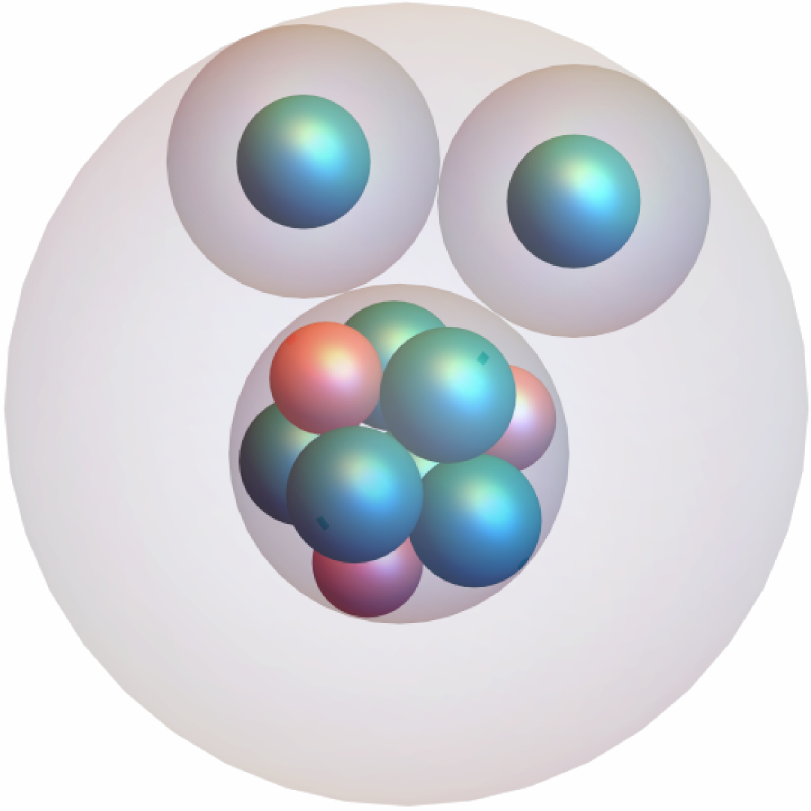}
	\includegraphics[width=0.24\textwidth]{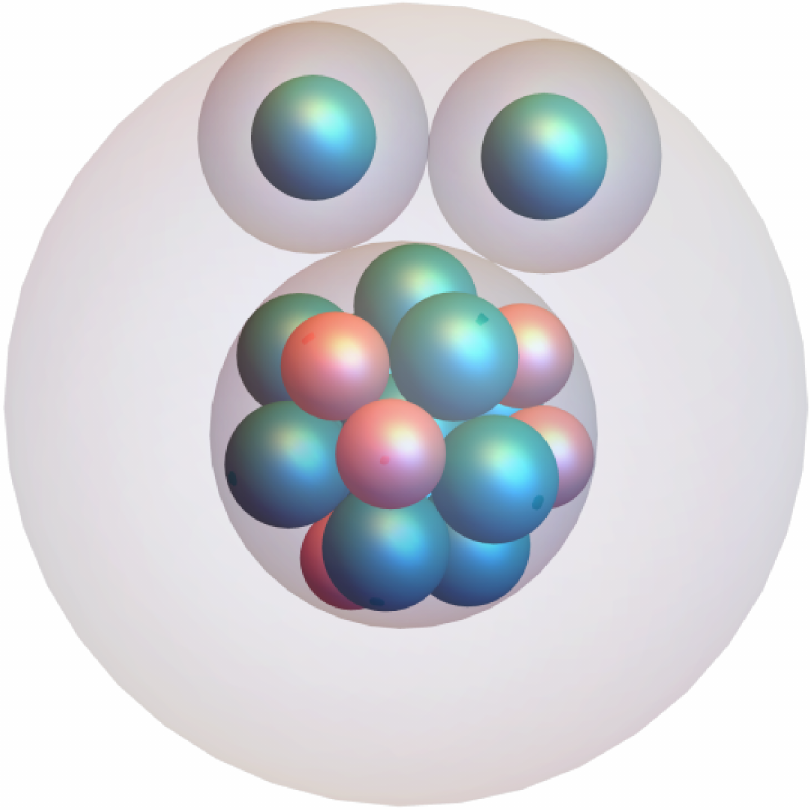}
	\includegraphics[width=0.24\textwidth]{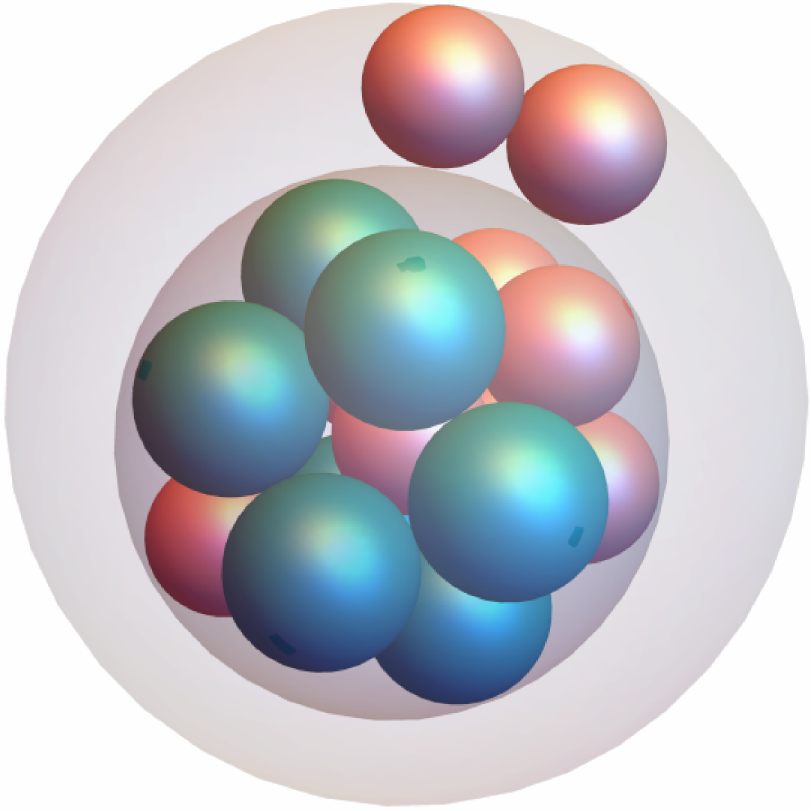}
	\caption{Visualizations of the three-cluster packing model for selected two-nucleon halo nuclei:  (a) the two-neutron halo nucleus $^{11}$Li represented as $^{9}$Li+$n$+$n$ with an added spatial offset $\Delta$ (“delta cloud”) illustrating the extended halo structure, (b) $^{19}$B as $^{17}$B+$n$+$n$ including the $\Delta$ extension, and (c) the two-proton halo $^{17}$Ne as $^{15}$O+$p$+$p$.}
	\label{fig:three-cluster-packing}
\end{figure*}

\begin{figure*}[ht!]
	\centering
	\includegraphics[width=0.25\textwidth]{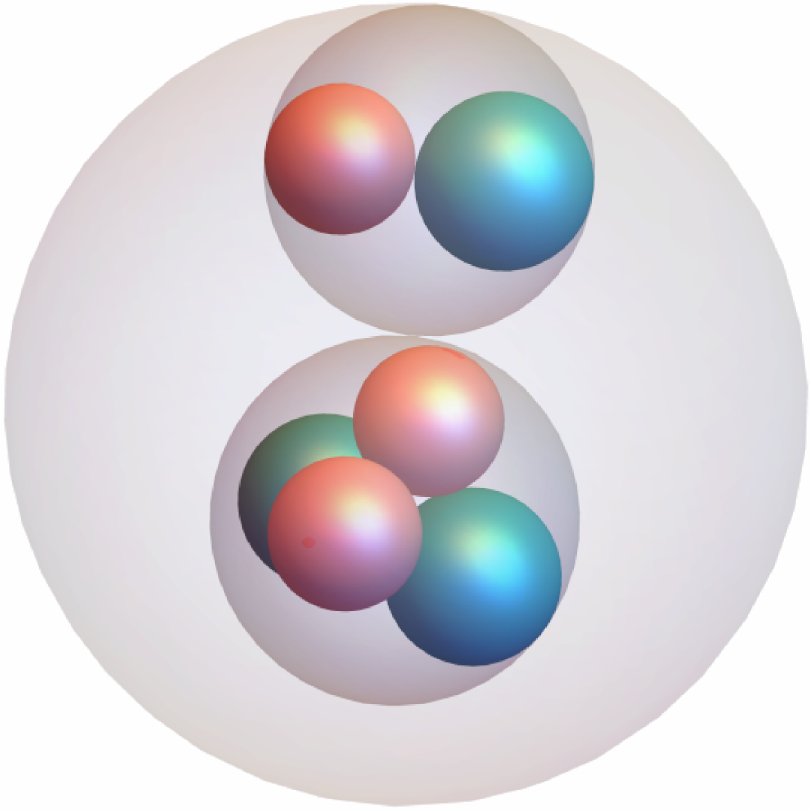}
	\includegraphics[width=0.25\textwidth]{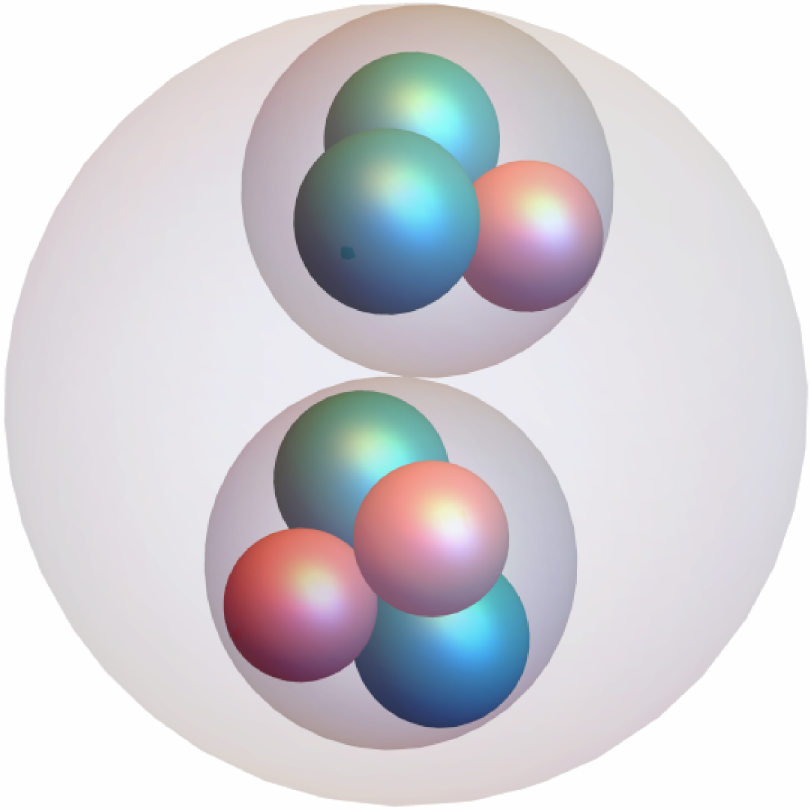}
	\includegraphics[width=0.25\textwidth]{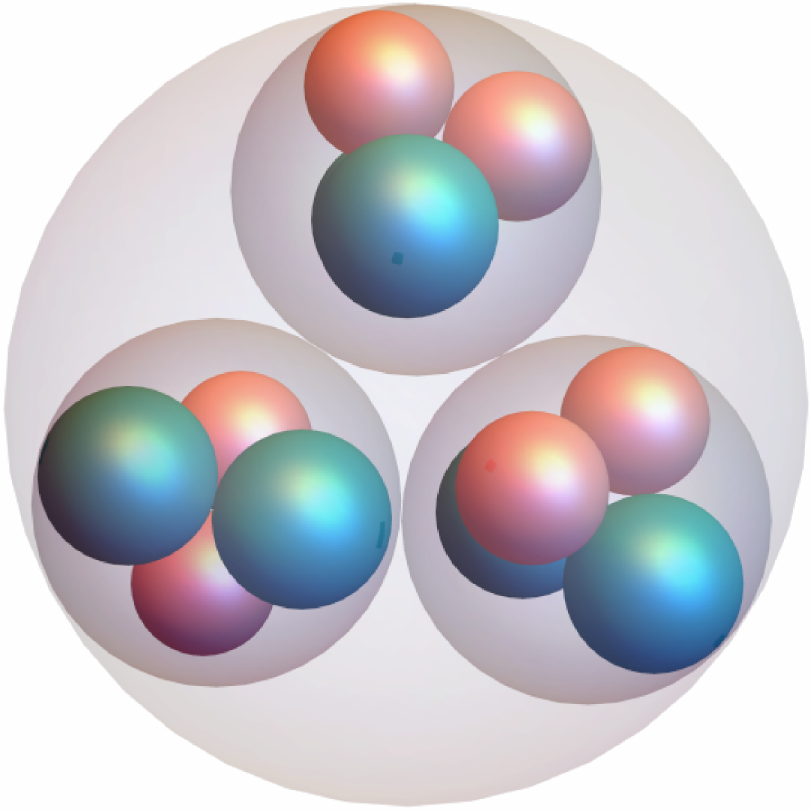}
	\caption{Visualizations of the sphere packing model of $^{6}$Li ($\alpha$+$d$), $^{7}$Li ($\alpha$+$t$), and $^{12}$C as three $\alpha$ particles.}
	\label{fig:alpha-cluster-packing}
\end{figure*}

The three-cluster packing model provides a geometric interpretation of two-nucleon halo systems, including both two-neutron and two-proton halos. In this approach, the nucleus is treated as a core surrounded by two weakly bound valence nucleons, which may couple either as a correlated dineutron (or diproton) cluster or as two independent nucleons. The spatial offset parameter~$\Delta$ determines the extent of the halo and is adjusted to reproduce the experimental core--valence and matter radii.

Table~\ref{tab:radii2} summarizes the calculated and experimental rms charge, matter, and core--valence radii, together with the neutron--neutron separation $r_{nn}$ and opening angle $\theta_{nn}$, for representative two-nucleon halo nuclei. Corresponding geometric configurations are illustrated in Figs.~\ref{fig:6He}, for $^{6}$He, and~\ref{fig:three-cluster-packing} for $^{11}$Li, $^{19}$B, and $^{17}$Ne.

For the two-neutron halo nucleus $^{6}$He, the three-cluster configuration ($^{4}$He+$n$+$n$) without an offset already yields reasonable agreement with experiment, giving $r_{m}^{\mathrm{rms}}=2.49$~fm compared with the measured values of 2.48--2.67~fm~\cite{Tan88,Ber07}. Introducing a modest offset of $\Delta=0.6$~fm extends the matter radius to 2.68~fm, while the core--valence distance $r_{cv}^{\mathrm{rms}}=3.90$~fm and neutron--neutron separation $r_{nn}=3.32$~fm are consistent with the experimental ranges 3.36--3.9~fm~\cite{Aum99,Sun21}. The corresponding opening angle $\theta_{nn}\approx51^{\circ}$ is in good agreement with the reported 52--56$^{\circ}$~\cite{Hag07,Sun21}, confirming that the model reproduces both the extended spatial distribution and the internal geometry of the halo.

The Borromean nucleus $^{11}$Li exhibits an even more diffuse structure. With $\Delta=1.0$~fm, the model reproduces a matter radius of $r_{m}^{\mathrm{rms}}=2.97$~fm, in good agreement with the experimental 2.78--3.12~fm~\cite{Tan88,Lia90}. The corresponding core--valence radius $r_{cv}^{\mathrm{rms}}=5.06$~fm closely matches the deduced experimental value of 5.01(32)~fm~\cite{Nak06} and 5.15(33)~fm~\cite{Hag07}, while the neutron--neutron separation $r_{nn}=4.12$~fm and opening angle $\theta_{nn}\approx51^{\circ}$ are consistent with the extracted geometry~\cite{Hag07,Nak06}. These results indicate that a relatively small $\Delta$ value reproduces the large spatial extension and correlated halo structure characteristic of $^{11}$Li.

For the heavier system $^{19}$B, a finite offset of $\Delta=0.9$~fm is required to match the experimental matter radius of $r_{m}^{\mathrm{rms}}=3.11(3)$~fm~\cite{Suz99}. The model predicts $r_{m}^{\mathrm{rms}}=3.14$~fm and $r_{cv}^{\mathrm{rms}}=5.70$~fm, in good agreement with the experimentally deduced core--valence radius of 5.75(11)~fm~\cite{Coo20}. 

The two-proton halo nucleus $^{17}$Ne is similarly well described by the three-cluster configuration ($^{15}$O+$p$+$p$). Without requiring any offset ($\Delta=0$), the model yields $r_{m}^{\mathrm{rms}}=2.80$~fm and $r_{cv}^{\mathrm{rms}}=4.34$~fm, consistent with the experimental matter radius of 2.75(7)~fm~\cite{Tan88}. 

Generally, the three-cluster packing model provides a coherent description of the geometric and structural properties of two-nucleon halo nuclei. The fitted $\Delta$ values systematically increase with decreasing two-nucleon separation energy, reflecting the progressive spatial decoupling of the valence nucleons from the core. The consistency between calculated and experimental radii, as well as between predicted and observed $r_{nn}$ and $\theta_{nn}$ values, demonstrates that the geometric offset parameter $\Delta$ effectively captures the essential configuration of weakly bound three-body halo systems.

\subsection{$\alpha$--Clustering } 
Although the main focus of this study is on weakly bound halo systems, the same geometric principles of \textsc{NUCLEI-PACK} extend naturally to $\alpha$--conjugate nuclei.
Figure~\ref{fig:alpha-cluster-packing} shows representative configurations of $^{6}$Li ($\alpha$+$d$), $^{7}$Li ($\alpha$+$t$), and $^{12}$C modeled as three $\alpha$ clusters, representative of the Hoyle-state geometry. Although quantitative radii are not evaluated here, these visualizations demonstrate that the packing framework reproduces the well-established spatial arrangements of $\alpha$ clusters in light nuclei directly from nucleon-level geometry. 

A detailed quantitative analysis of $\alpha$--cluster states, including excited configurations and comparisons to experimental radii, will be presented in a forthcoming study.

\section{Conclusion}
In this work, a semi-classical packing model has been developed and applied to describe the geometric structure of light and exotic nuclei. Within this framework, nucleons are treated as spatially localized spheres whose optimal packing defines the equilibrium configuration of the system. By explicitly separating compact cores from valence nucleons or clusters, the model reproduces the geometry and spatial extension of the halo systems.

For one-nucleon halos, the two-cluster extension of the model reproduces the experimental charge and matter radii of $^{11}$Be, $^{15}$C, $^{19}$C, and $^{8}$B with remarkable accuracy. The fitted offset parameter~$\Delta$ exhibits a clear inverse correlation with the one-nucleon separation energy, demonstrating that more weakly bound systems require larger geometric displacements between the core and the valence nucleon to reproduce the observed rms radii. This simple geometric parameter effectively captures the core--valence characteristics.

The three-cluster version of the model successfully extends this concept to two-nucleon halos. Calculations for $^{6}$He, $^{11}$Li, $^{19}$B, and $^{17}$Ne reproduce the magnitudes of the experimental matter radii, the core--valence distance, the  neutron--neutron (or proton--proton) separations and the opening angles inferred from measurements. The model naturally captures the transition from compact to highly diffuse geometries, reflecting the underlying reduction in binding energy and the correlated motion of valence nucleons in Borromean systems.

Beyond halo phenomena, the same geometric formalism can qualitatively describe $\alpha$--cluster nuclei such as $^{6}$Li, $^{7}$Li, and $^{12}$C, including the iconic Hoyle state. These results demonstrate the versatility of the packing framework as a unified geometric picture of clustering and halo formation in light nuclei.

Future work will aim to incorporate microscopic quantum effects, including spin--orbit coupling and pairing correlations, to refine the predictive power of the model. Coupling the packing algorithm with machine learning and optimization techniques will enable large-scale surveys of nuclear geometries across the chart of nuclides. Ultimately, the geometric packing approach provides a bridge between macroscopic intuition and microscopic structure, offering a transparent and computationally efficient framework to understanding the emergence of clustering and halo phenomena in nuclei.


\end{document}